\documentclass[aps,showpacs,twocolumn,amsmath,amssymb]{revtex4}
\usepackage{graphicx}
\usepackage{dcolumn}
\usepackage{bm}
\begin{document}

\title{The symmetry of the Kuramoto system and the essence of the cluster synchronization}

\author{Guihua Tian$^{1,2}$,\ \ Songhua Hu$^{1,2,3}$}\email{tgh-2000@263.net, husonghua@126.com }
 \affiliation{$^1$School of Science, Beijing
University of Posts And Telecommunications. Beijing 100876, China.}
 \affiliation{$^2$State Key Laboratory of Information Photonics and Optical
Communications, \\ Beijing University of Posts And Telecommunications.
 Beijing 100876, China.}
 \affiliation{$^3$ School of Electronic and Information Engineering, \\ North China Institute of Science and Technology,Yanjiao 065201, China}
\begin{abstract}
    The cluster synchronization (CS) is a very important characteristic for the higher harmonic coupling  Kuramoto system. A novel method from the symmetry transformation is provided, and it gives CS a profoundly mathematical explanation and clear physical annotation. Detailed numerical studies for the order parameters in various conditions confirm the theoretical predictions from this new view of the symmetry transformation. The work is very beneficial to the  further study on CS in various systems.
\end{abstract}
\maketitle

As the simplest  and the most celebrated one, the Kuramoto model  captures the main property of the collective synchronization, and is applied in many physical, biological and social systems, including electrochemical oscillators, Josephson junction arrays,  cardiac pacemaker cells, circadian rhythms in mammals, network structure and neural network\cite{kura}-\cite{gupta}.

In the large-N limit, the Kuramoto model revealed the second continuous transition at the critical coupling strength $K_c$. Many generalizations of the Kuramoto model have been investigated. Including the large inertia  in the generalized Kuramoto model, the transition from the incoherent state to the collective synchronization became of the first order\cite{gupta,tana}. Noise also can push the incoherent stationary state to become stable\cite{aceb2,aceb3}.
When the universal coupling strength $K$ becomes oscillator-dependent and correlated with the frequency, the explosive synchronization (ES) appears\cite{wang}-\cite{juses}. ES is also an abrupt, of the first-like phase transformation. The identical oscillators with the nonlocal coupling strength will give rise to the new  chimera state, which is the combination of the coherent state and the incoherent state for the identical oscillators\cite{abram}-\cite{ping}.

All the above examples are of the first harmonic coupling as $H(\theta_j-\theta_i)=K_{ij}\sin (\theta_j-\theta_i) $. Whenever the higher globally coupling harmonic term is introduced, interesting phenomena appear, like the cluster synchronization (CS) or multi-entrainment, and switching of the oscillators between different clusters with
the external force  \cite{ott}-\cite{koma}. The higher harmonic coupling is dominating in $\phi$-Josephson junction \cite{gold,gold2}, in the electrochemical oscillators in higher voltage\cite{kiss,kiss2,ott}, in neuronal networks with learning and network adaption\cite{seli}-\cite{niyo similar}. CS is the most outstanding feature of the Kuramoto model with higher order harmonic coupling.

CS or multi-entrainment has been investigated by the method of self-consistent approach in Refs.\cite{ott},\cite{koma} ,\cite{seli}-\cite{ golo}. Neural network actually studied the combination of the first and second harmonic couplings in the generalized Kuramoto model\cite{seli}-\cite{golo}, which is also treated  in Ref.\cite{koma}, \cite{koma2}.
In the $N$ identical oscillators'case, the symmetry viewpoint is applied and CS of the two groups of $m$ and $N-m$ oscillators   is connected with their symmetry groups of the dynamics $S_m\times S_{N-m}$ \cite{Ashwin}, \cite{ban}. The symmetry group $S_N$ is only suited for the identical oscillators in the Kuramoto model. However, it is still very difficult to obtain clear analytical results by the self-consistent approach and detailed understanding of the stability of the asynchronous states is still missing \cite{koma, koma2}.

In order to overcome the shortcoming,
the density function for the second harmonic coupling case is decomposed into the symmetric and asymmetric parts in Ref.\cite{ott} , and the Ott-Antonsen (OA) mechanism is utilized to analyze the symmetric case. Ott-Antonsen  ansatz (reduction mechanism) is  powerful in  obtaining the analytical understanding of the Kuramoto model\cite{otta} and has been applied in generalized Kuramoto ones, like the forced Kuramoto \cite{child}, the bimodal frequency distribution\cite{mart,pazo} and the second harmonic coupling\cite{ott}.
The asymmetric clustering is also showed in Ref.\cite{ott} very sensitive to the non-uniform initial condition. Even the OA ansatz can not give a clear physical understanding of CS thanks to unavailability  in obtaining the analytic form for the intricate  asymmetry density function \cite{ott}.

Here we will investigate the higher-harmonic coupling Kuramoto model from the point of symmetry, and provide a group transformation, which is completely different from that of  $S_N$ group. Then we give CS a thoroughly novel interpretation.

In the original Kuramoto model \begin{eqnarray}
&&\dot{\theta_n}=\omega_n+\frac{1}{N}\sum_{j=1}^{N}K\sin (\theta_j-\theta_n) \label{Kuramoto ori},
\end{eqnarray}
the coupling strength $K>0$ is assumed.
The higher harmonic coupling of the generalized Kuramoto model is
\begin{eqnarray}
&&\dot{\theta_n}=\bar{\omega}_n+\frac{1}{N}\sum_{j=1}^{N}\bar{K}_m\sin m(\theta_j-\theta_n) \label{Kuramoto mth}.
\end{eqnarray}
The order parameter is defined as
\begin{eqnarray}
&&re^{i\Psi}=\frac{1}{N}\sum_{j=1}^{N} e^{i\theta_j}.\label{order in theta}
\end{eqnarray}
Ref.\cite{ott} shows that the critical parameter $\Bar{K}_m^c$ relating with the corresponding $m$-th order parameter is the same $\Bar{K}_m^c=2\Delta$ for all integers $m$, where $2\Delta$ is  the width of the original Lorentz frequency distribution in Ref.\cite{ott}.  In the case of small strength $\Bar{K}_m<\Bar{K}_m^c$, the term $\omega_n$ dominates the change of the phase $\theta_n$ and the whole phase system is in the incoherent  state. Whenever $\Bar{K}_m$ exceeds $\Bar{K}_m^c$, the second terms in Eq.(\ref{Kuramoto mth}) predominate and CS emerges\cite{ott}.

Why the critical coupling $\Bar{K}_m^c$ is the same for all different integers $m$? Is it only a coincidence or is there some underlying reason? The study in the letter shows that  the symmetry that the Kuramoto system keeps is responsible for.

By introduction of the transformation
\begin{eqnarray}
\phi=m\theta,\ \omega_n=m\bar{\omega}_n,\ K=m\Bar{K}_m,\  \label{trans}
\end{eqnarray}
Eq.(\ref{Kuramoto mth}) takes the form
\begin{eqnarray}
&&\dot{\phi_n}=\omega_n+\frac{1}{N}\sum_{j=1}^{N}K\sin (\phi_j-\phi_n) \label{Kuramoto 1-nth},
\end{eqnarray}
which is the same as Eq. (\ref{Kuramoto ori}) of the standard Kuramoto model\footnote{After completing our work, we notice Ref.\cite{niyo similar} has a similar transformation for $m=2$ case in the fast study model.}.

The generalized order parameters is defined as $R_{1}=\frac{1}{N}\sum_{j=1}^{N} e^{i\phi_j} =|R_1| e^{i\varphi_1}$,
and Eq.(\ref{Kuramoto 1-nth})   becomes
$\dot{\phi_n}=\bar{\omega}_n+\frac{K}{2i}(R_{1}e^{-i\phi_n}-  R^*_{1}e^{i\phi_n} )$.
In the limit $N\rightarrow \infty$, the density function $f(\phi,\omega,t)$ is introduced to
describe the distribution of the phases at a given frequency and satisfies the continuous equations
\begin{equation}
\partial_tf+\partial_{\phi}\bigg[\bigg(\omega+\frac{K}{2i}\bigg(R_{1}e^{-i\phi}-  R^*_{1}e^{i\phi} \bigg)\bigg)f\bigg]=0.
\end{equation}
The solutions to Eq.(\ref{Kuramoto mth}) could be obtained from the ones to Eq.(\ref{Kuramoto 1-nth}), where OA mechanism could be utilized when the distribution of the natural frequency is the Lorentz's one \cite{otta}.

The density distribution function $f(\phi,\omega,t)$ is a periodic function satisfying $f(\phi+2\pi,\omega), t)=f(\phi,\omega,t)$, and in the case $m=2$ it corresponds to the symmetric one in Ref\cite{ott}.
It is easy to see that in the stationary state $f(\phi,\omega,t)=f(\phi,\omega,t)|_{t=\infty}$ the system is partially synchronic whenever
$K>2\Delta$ for the Lorentz distribution of the natural frequency of the phases with $\Delta$ its width\cite{ott}, \cite{otta}, \footnote{The critical coupling strength for $K_m$ is the same $2\Delta$ for the Lorentz frequency distribution, and one should not use the relation $K=m\bar{K}_m$ to deduce the critical strength for Eq.(\ref{Kuramoto mth}).  The correct deduction is from the fact that the critical strength both for oscillators in $\theta$ or in $\phi$ is obtained from the density function $f(\phi,\omega,t)$.  The conclusion is the  critical strengthes for $K$ and $K_m$ are the same  for both Lorentz frequency distribution and   other frequency distributions.}. Hence the same critical parameter $K_c $ is realized for all $m$-th order parameters. Combination of the symmetry transformation (\ref{trans}), $f(\phi,\omega,t)$ can completely determine the evolution of the dynamics and this is the key point in the latter to study CS.

 We apply both the transformation (\ref{trans}) and the distribution function $f(\phi,\omega,t)$ with its periodic property to investigate the order parameters for the oscillators in several special cases, and make the corresponding predictions on the order parameters. Then we  integrate Eq.(\ref{Kuramoto mth}) directly for these cases and obtain the corresponding numerical order parameters directly from the numerical integration. The later numerical results confirm the former prediction. The details are divided into five groups in the following.

(a)If the initial oscillators' phases are uniformly distributed in $(0, 2\pi)$, together with  the transformations (\ref{trans}), the order parameter $r$ in Eq.(\ref{order in theta}) in the large $N$ limit turns out as  \begin{eqnarray}
re^{i\Psi}&=&\frac1m \int_{-\infty}^{+\infty}\int_0^{2m\pi}f(\phi,\omega,t)e^{i\frac{\phi}m}d\omega d\phi =0,\label{cluster r}
\end{eqnarray}
where the upper integral number is $2m\pi$ due to the transformations (\ref{trans}). The symmetry property of the distribution function $f(\phi+2\pi,\omega), t)=f(\phi,\omega, t)$ makes the order parameter
$r=0,$
no matter what a great coupling strength is applied. This is actually the manifestation of the cluster property of the higher harmonic coupling. The $m$-term harmonic coupling will give rise to the corresponding $m$ clusters, and   the phase oscillators in one cluster behave completely the same way as those in another cluster.
The order parameter $r$ can only take the zero value, which is a typical manifestations for  the cluster phenomena of  the higher harmonic coupling in the  system.
\begin{center}
\begin{figure}[ht]
\begin{tabular}{cc}
\includegraphics[height=0.10\textheight, width=0.22\textwidth]{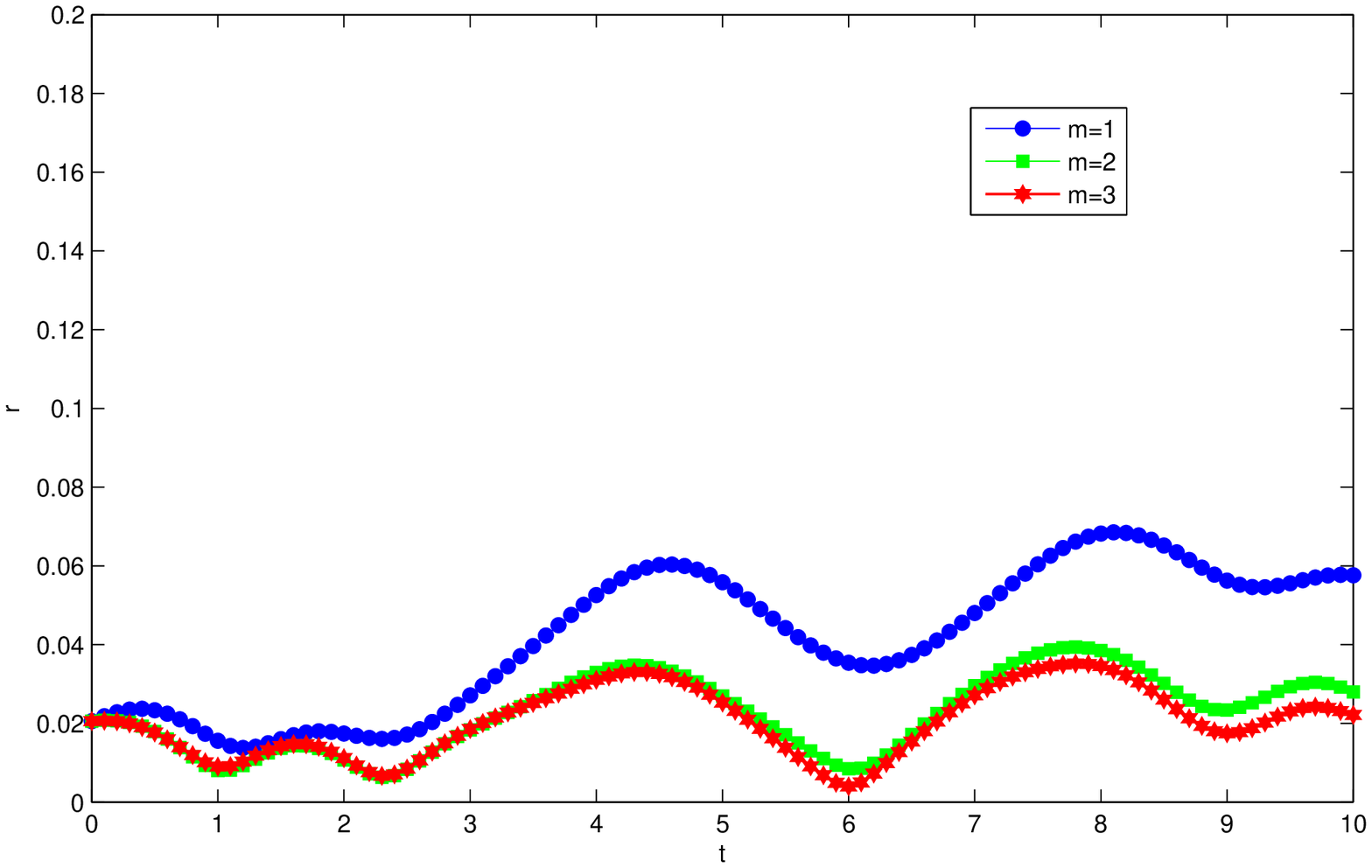}&
\includegraphics[height=0.10\textheight, width=0.22\textwidth]{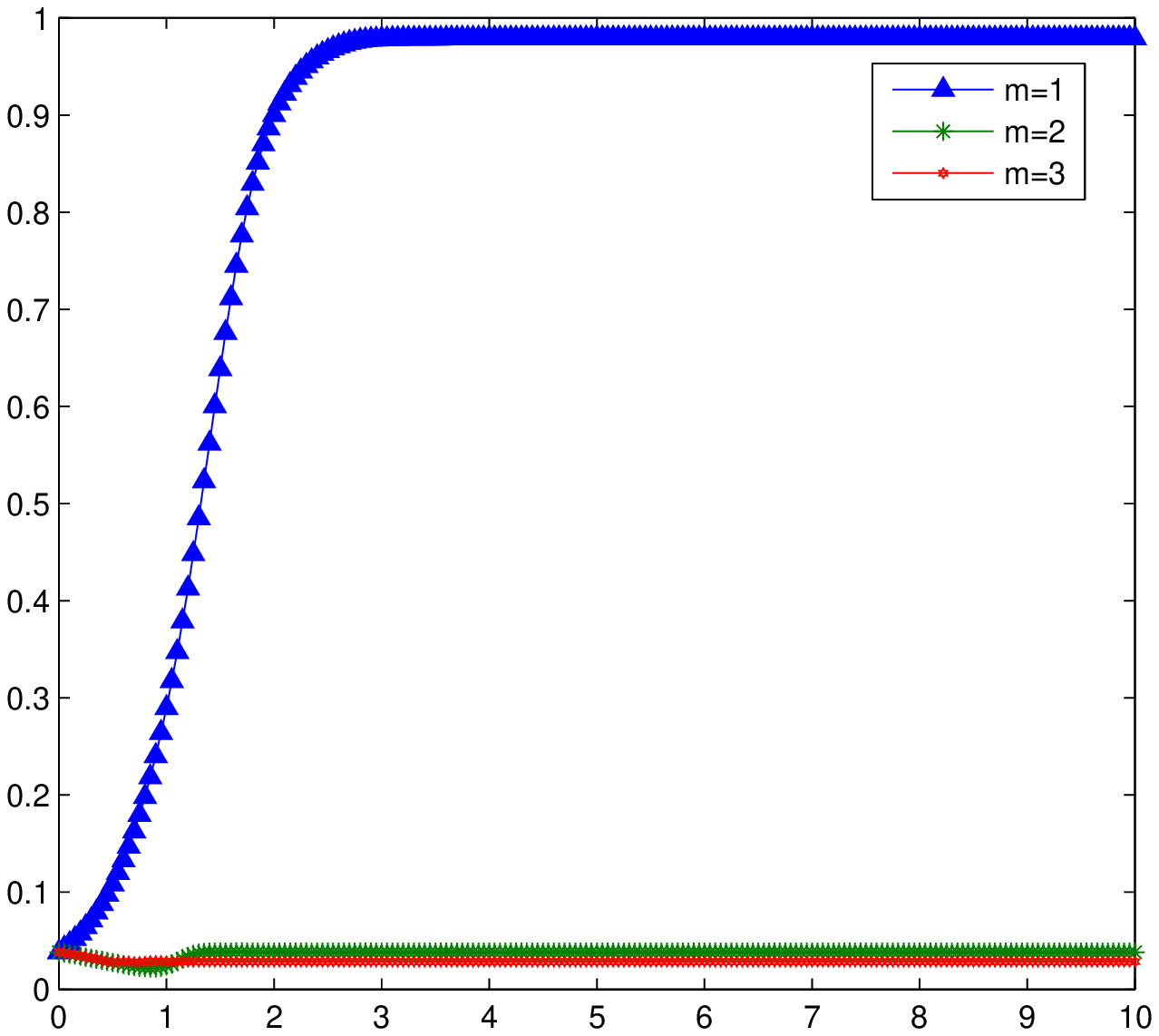}
\end{tabular}
\caption{Diagrams $r(t)$ of initially uniform distribution of the phases  with the same coupling strength $K=1,\ K<K_c$ in the left panel indicating the incoherent state and $K=5,\ K>K_c$ in the right panel showing CS for $m=2,\ 3$. Notice that the green lines  are shielded by the red lines.  The number of the oscillators is $N$ in all the figures}. \label{fig1}
\end{figure}
\end{center}
The  numerical studies in Fig.\ref{fig1} confirm the above ideas of $r\approx 0$ with the uniform initial distribution in $(0,2\pi)$ of the phases and different coupling strengths and different higher-term couplings. When $K<K_c$, all parameters $r$ in the three cases in Fig.\ref{fig1} is similar, and the oscillators are incoherent. Of course, the symmetry properties in Eq.(\ref{cluster r}) will force parameters $r$ in $m\ge 2$ will be more smaller than that in $m=1$ case. When the coupling strengths $K$ surpass $K_c$, $r\approx 1$ is obtained for usual Kuramoto model (m=1), and $r\approx 0$ stands out in $m\ge2$  in Fig.\ref{fig1}, which indicates the formation of the clustering synchronization. Fig\ref{fig2} further gives clustering synchronization for the case of $K=5, \ m=2,\ m=3$ by the phases' position in the circles  in the middle and the left panels.
\begin{center}
\begin{figure}[ht]
\begin{tabular}{ccc}
\includegraphics[height=0.10\textheight, width=0.15\textwidth]{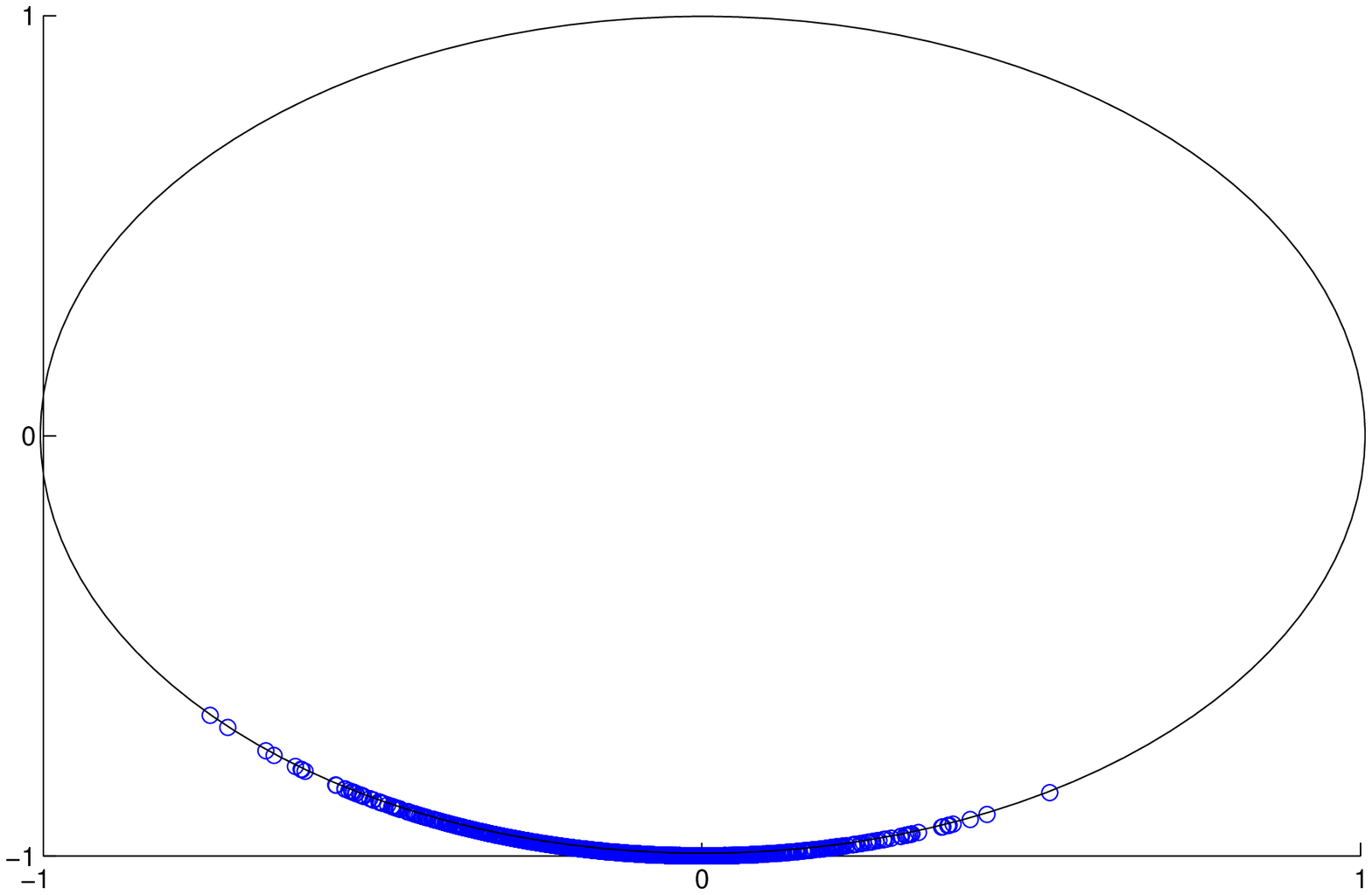}&
\includegraphics[height=0.10\textheight, width=0.15\textwidth]{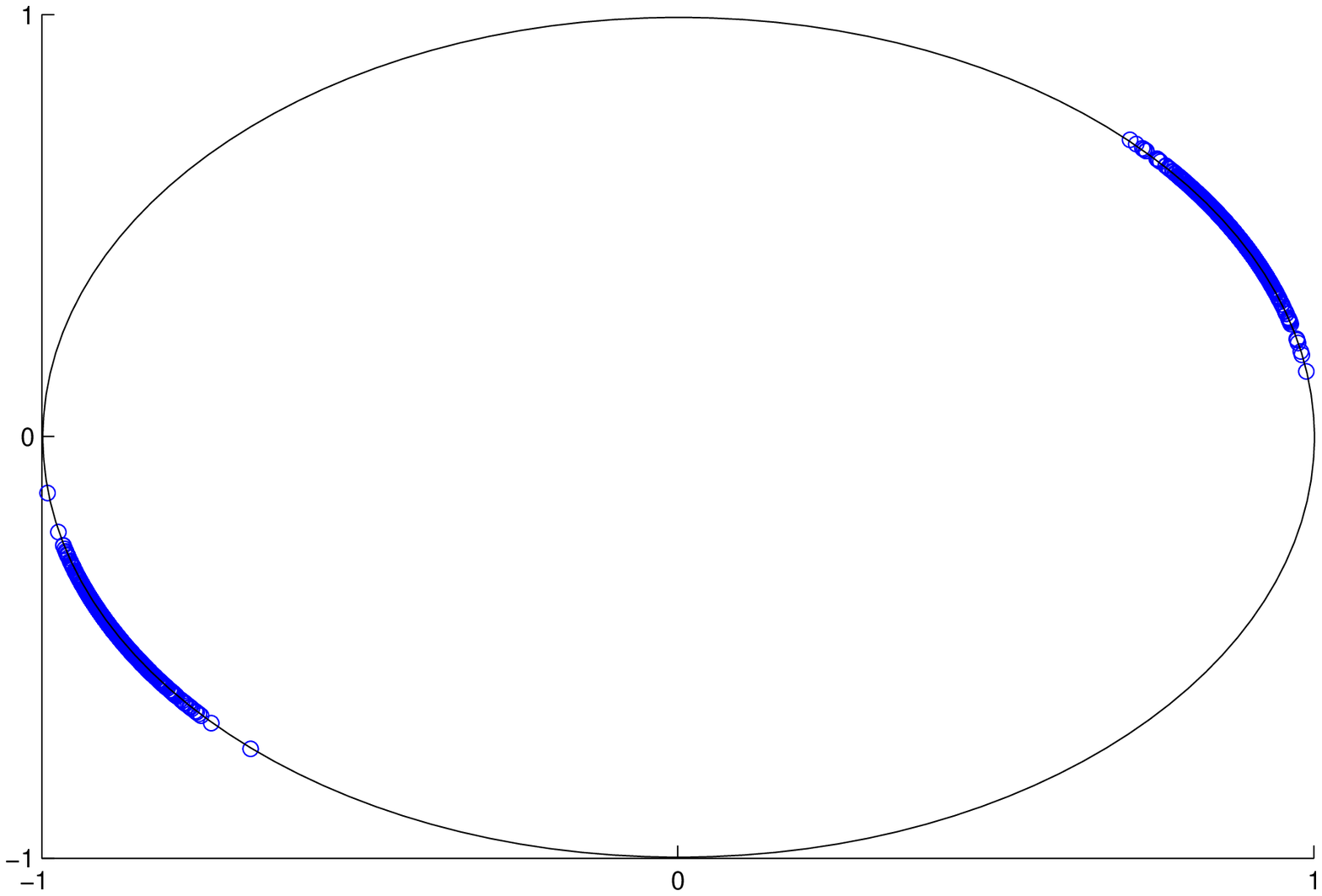}&
\includegraphics[height=0.10\textheight, width=0.15\textwidth]{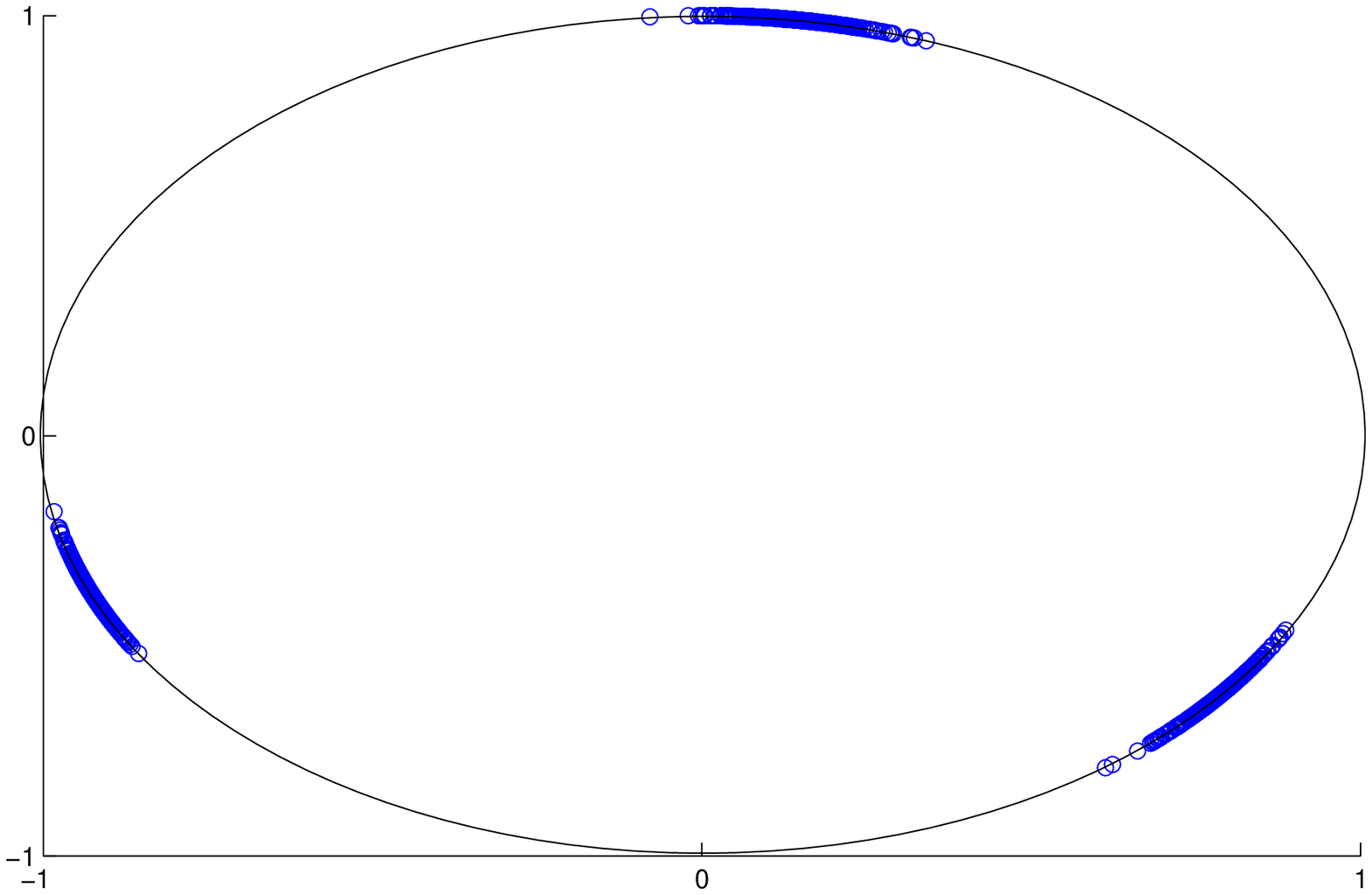}
\end{tabular}
\caption{The cluster synchronization of the phases in the circles corresponding the parameters in the second panel in Fig.1  with the same parameters $K=5$ but different  $m=1,2,3$. There are $1000$ oscillators and their positions are indicated by the small blue circles in the big circles.}\label{fig2}
\end{figure}
\end{center}
Furthermore, the more higher harmonic coupling is applied, the more clusters appear and the smaller section of the whole range $2\pi$ every cluster occupies. Therefore, very higher harmonic coupling of the oscillators will produce psudo-synchronization where the large number of the clusters will globally behave the same way with the oscillators within each very small  cluster being any state.

In each cluster section, the phase oscillators may stay in the incoherent state or partially synchronic state, even in synchronization state depending on whether or not $K>K_c$. The order parameter for the phase oscillators in each cluster is the same and is $
r^s_1=|\int_{-\infty}^{+\infty}\int_{-\pi+\phi_f}^{\pi+\phi_f}f(\phi,\omega,t)e^{i\frac{\phi}m}d\omega d\phi | \ne 0$
for the first  cluster, where $\phi_f $ is the center of the first cluster.
The order parameters $r^s_1$ and $r$ are different from each other, but the critical coupling strengthes for them are the same.  The cause is  that both $r^s_1$ and $r$ are determined by the same distribution function $f(\phi,\omega,t)$, which decides the critical strength $K_c$.

(b) When the initial phases distribution $(0, A)$ is narrowed and much smaller than $\frac{2\pi}m$, and the coupling strength is stronger than the critical $K_c$, and with every term in the second part $\frac{1}{N}\sum_{j=1}^{N}\bar{K}_m\sin m(\theta_j-\theta_n)$ in Eq.(\ref{Kuramoto mth}) has the same effect as the corresponding term in ordinary Kuramotto (\ref{Kuramoto ori}), and they together
dominate over the first part and attract all oscillators to the synchronic state. So, the initial synchronized state in one cluster will remain synchronized all the time, just like in the ordinary Kuramoto model and the order parameter $r=|\int_{-\infty}^{+\infty}\int_0^{2\pi}f(\phi,\omega,t)e^{i\frac{\phi}m}d\omega d\phi | \approx r_1^s
\approx 1 $ is realized. In this case, almost all oscillators are synchronized in the only one cluster, and no other cluster synchronization exists. See the phases in the circles  in the second ($A=0.75\pi,\ m=2$) and third ($A=0.8\pi,\ m=2$)  panels in Fig.\ref{fig3} for details. Note there are several phases are left opposite to the synchronized one in the two panels.
\begin{widetext}
\begin{center}
\begin{figure}[ht]
\begin{tabular}{ccccc}
\includegraphics[height=0.10\textheight, width=0.2\textwidth]{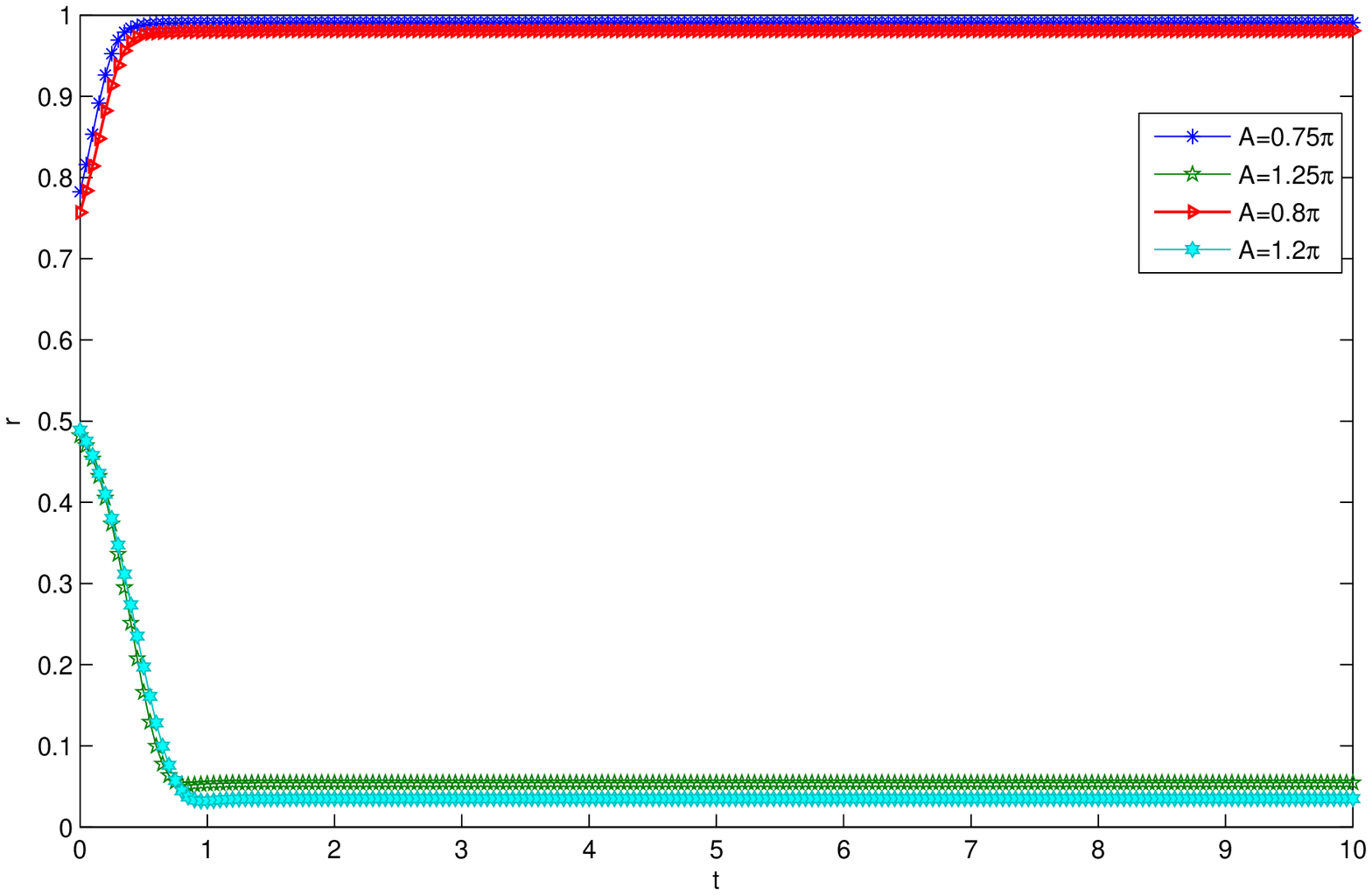}&
\includegraphics[height=0.10\textheight, width=0.18\textwidth]{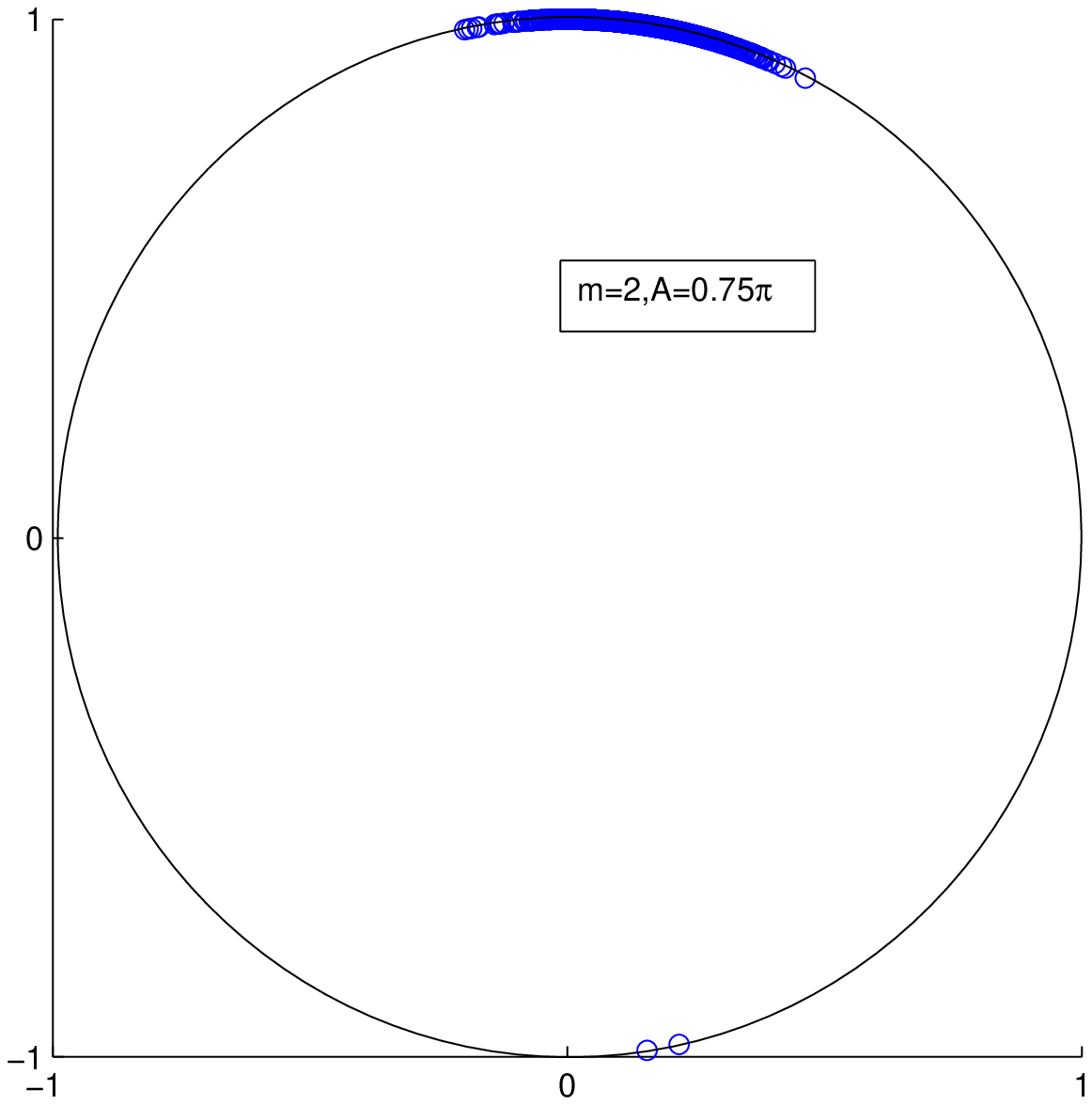}&
\includegraphics[height=0.10\textheight, width=0.18\textwidth]{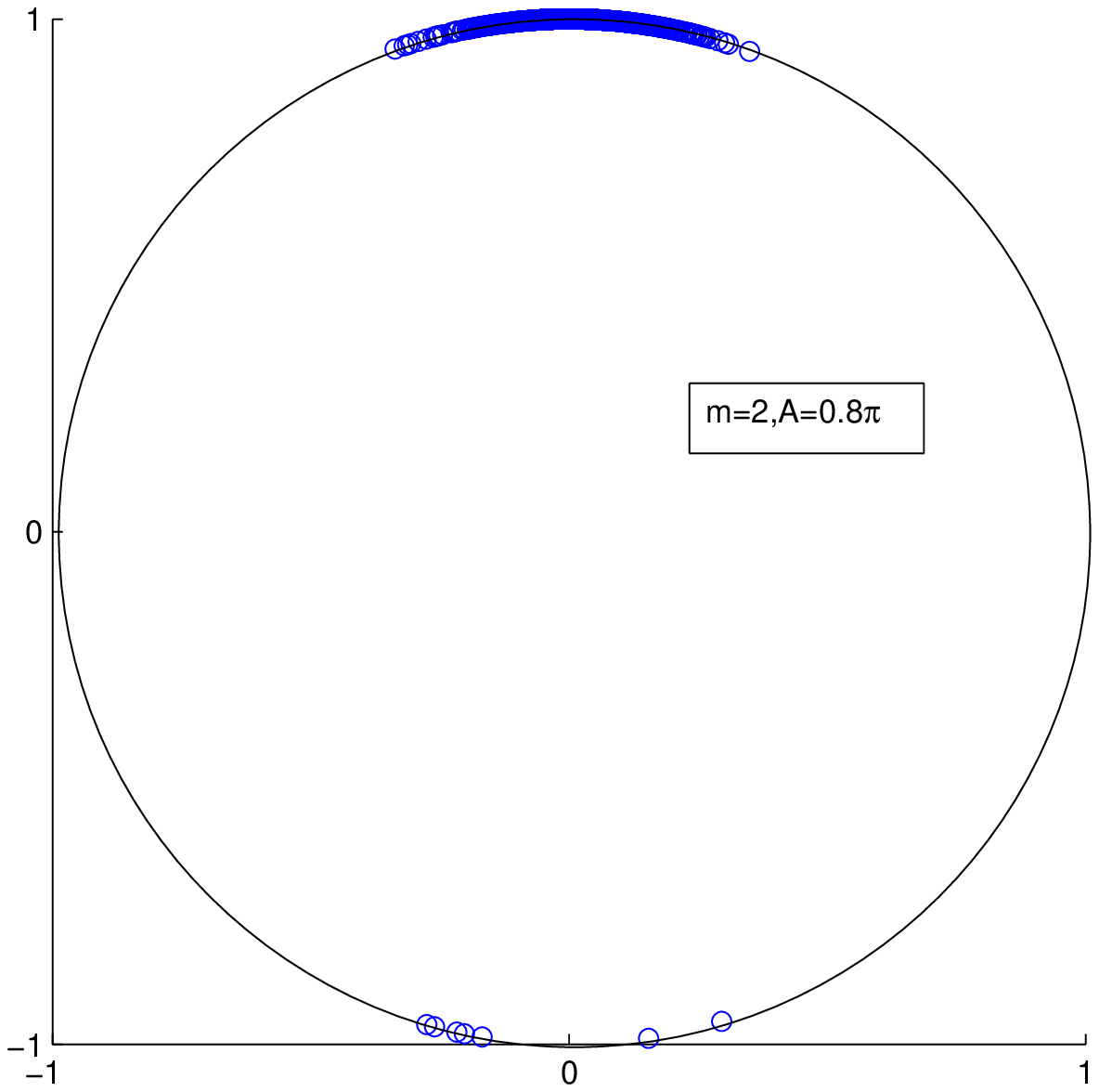}&
\includegraphics[height=0.10\textheight, width=0.18\textwidth]{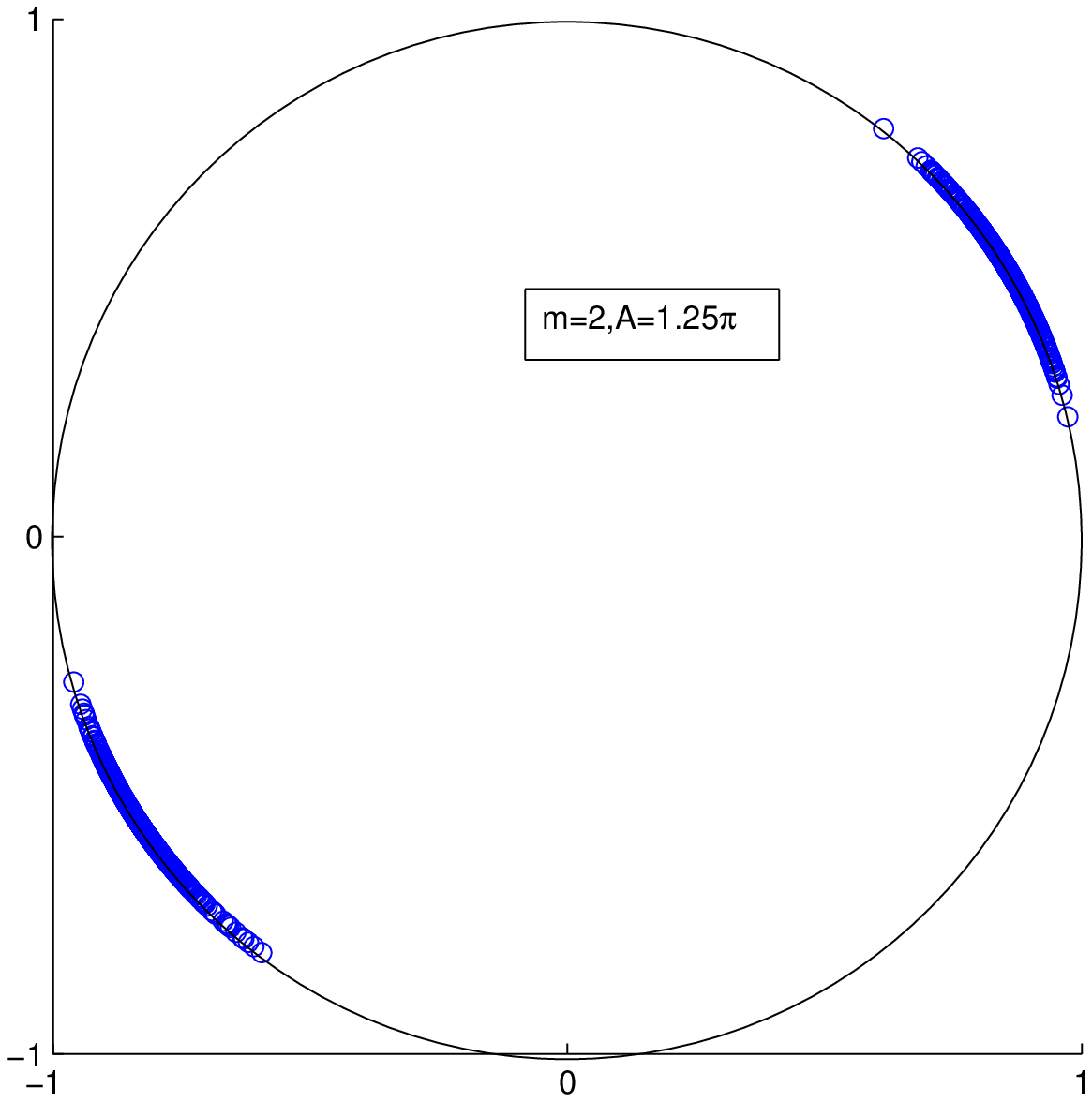}&
\includegraphics[height=0.10\textheight, width=0.18\textwidth]{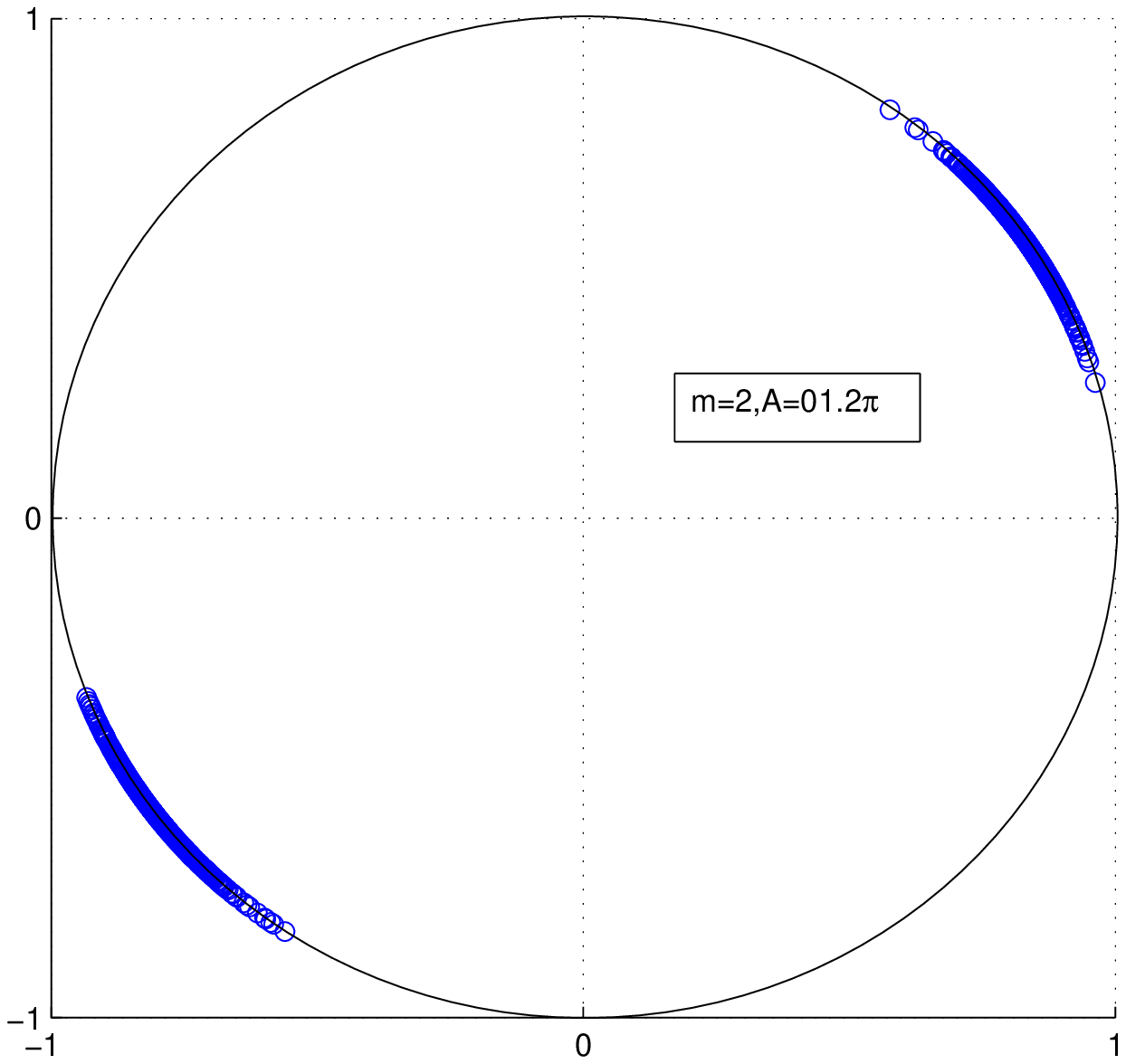}
\end{tabular}
\caption{Schematic diagrams $r(t)$ of initial uniform phases distribution in $(0,A)$   with the same parameters $K=5,\ m=2$ but different  $A$.} \label{fig3}
\end{figure}
\end{center}
\end{widetext}
(c)When $A$ of the initial distribution $(0, A)$  goes beyond   $\frac{2(m-1)\pi}m$ but less than $2\pi$, there approximately emerges $m$  cluster synchronization. As the order parameter $r\approx 0$ is achieved, which is  shown in  Fig.\ref{fig3} with $A=1.2\pi,\ 1.25\pi$ in the case of $m=2,\ K=5$ for the Gauss frequency distribution with $K_c<2$. In this case, the initial range of $\phi=m\theta$ exceeds $2(m-1)\pi$, and guarantee the validation of the transformation (\ref{trans}) and Eq.(\ref{cluster r}). So CS is again  connected with the symmetry of the Kuramoto model (\ref{Kuramoto mth}). The fourth ($A=1.25\pi,\ m=2$) and fifth ($A=1.2\pi,\ m=2$) panels  in Fig.\ref{fig3} agree with this analysis.

\begin{center}
\begin{figure}[ht]
\begin{tabular}{cc}
\includegraphics[height=0.10\textheight, width=0.22\textwidth]{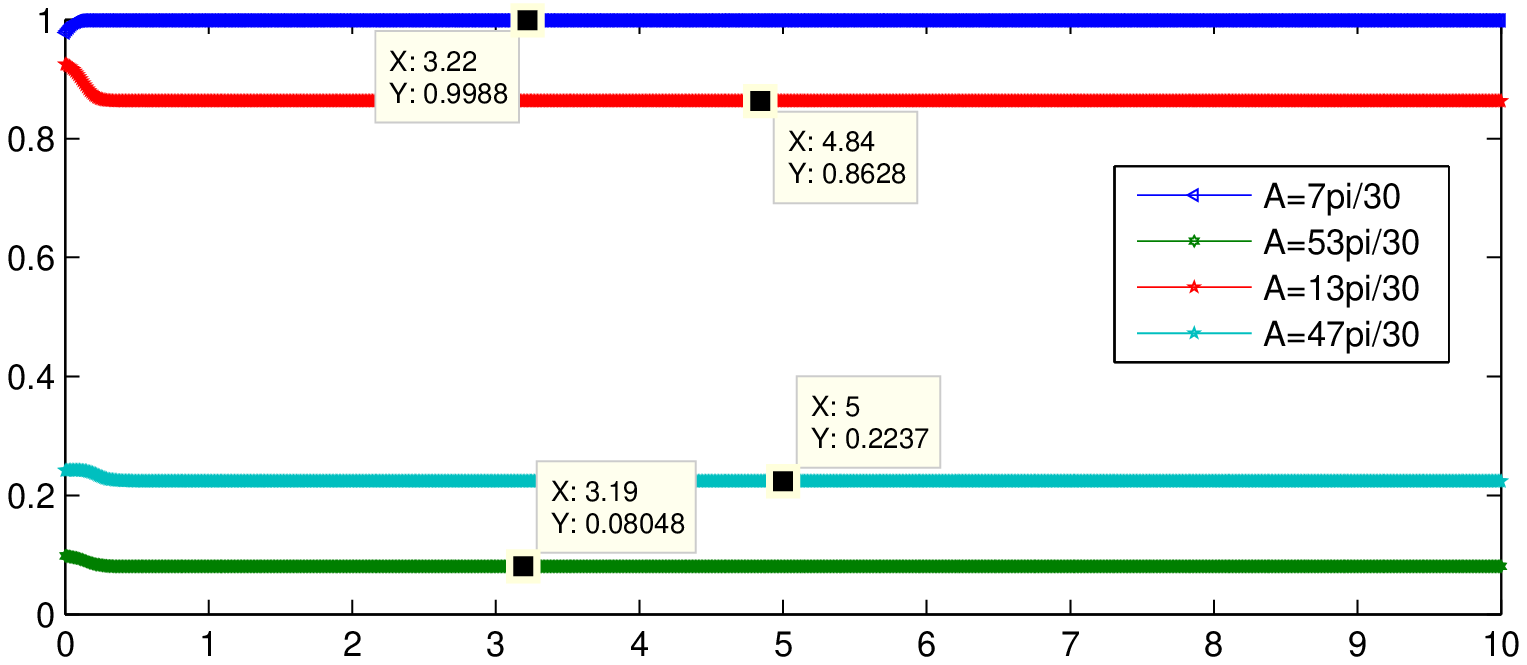}&
\includegraphics[height=0.10\textheight, width=0.22\textwidth]{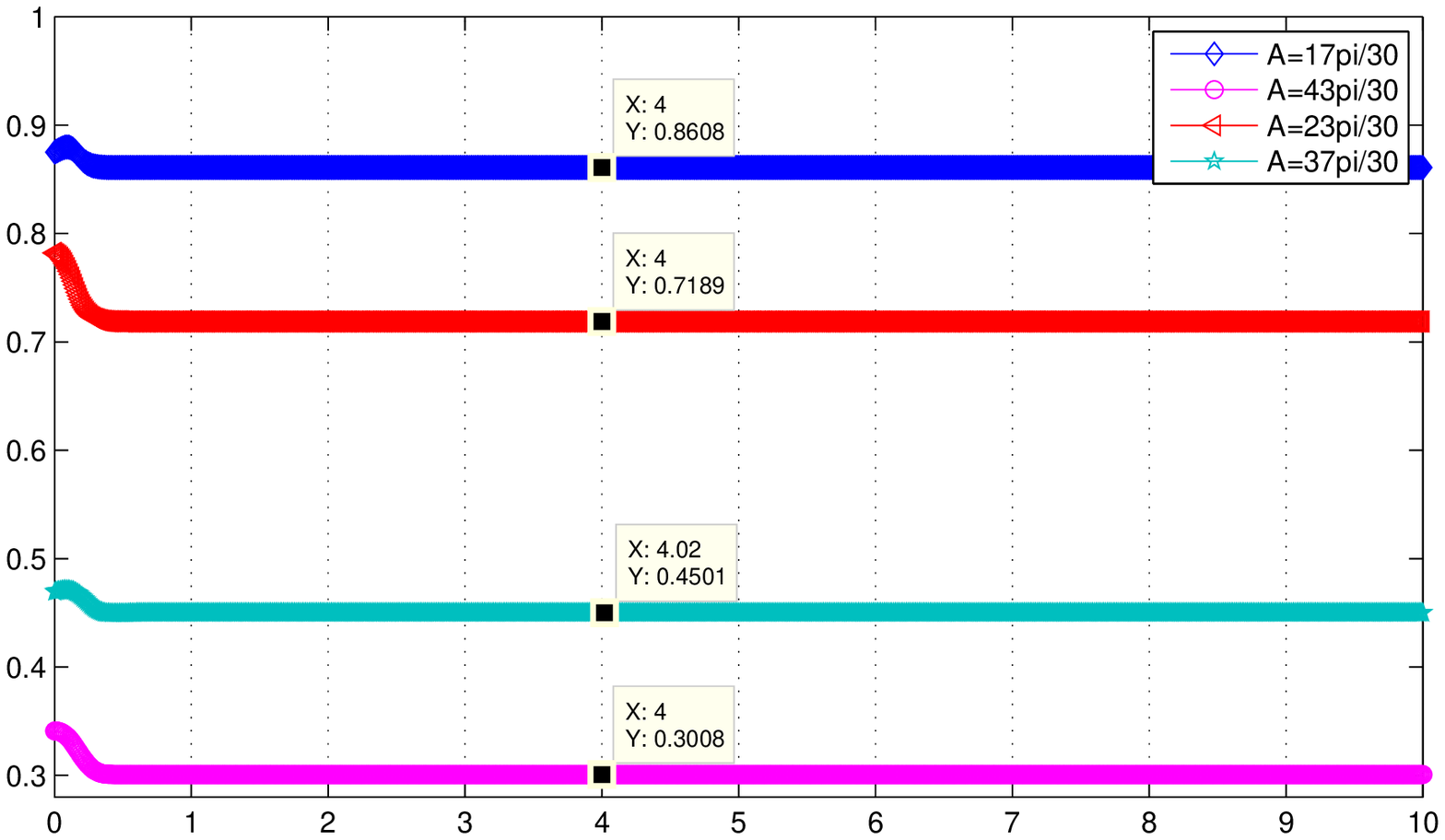}
\end{tabular}
\caption{Detailed diagrams $r(t)$ on the  initial uniform phases distribution in $(0, A)$   with the same parameters $K=5,\ m=6$ but different  $A$. }\label{fig4}
\end{figure}
\end{center}

(d)When $A$ of the initial distribution $(0, A)$  lies in $\frac{2(n-1)\pi}m$ to $\frac{2n\pi}m$ with   $n< m$ and $m$ greater than $2$, the partial CS appears indicated by the order parameter $r$ approaches  neither $0$ nor $1$.  The order parameter takes the form \begin{eqnarray}r(A)= |\int_{-\infty}^{+\infty}\int_0^{mA}f(\phi,\omega,t)e^{i\frac{\phi}m}d\omega d\phi |,\label{partial order}\end{eqnarray} Generally, in the stationary state, the upper bound $mA$ could be replaced by $2n\pi$ in most cases.
 Roughly, $r(A)\approx |\frac1{n}(\sum_{j=0}^{j=n}e^{i\frac{2j\pi}m})r_1^s|$, where $r_1^s$ is the first cluster's order parameter, and is near $1$ in the case $K$ being much large than $K_c$.
The  numerical studies  in Fig.\ref{fig4} crudely illustrate  the feature for $r(A)$.   Define $r(n)=|\int_{-\infty}^{+\infty}\int_0^{2n\pi}f(\phi,\omega,t)e^{i\frac{\phi}m}d\omega d\phi |$, it is easy to see that $r(n)<r(A)<r(n-1)$. For example, $r(n)$ takes the following numerical values $r(1)=1,\ r(2)\approx 0.8666,\ r(3)\approx 0.6777,\ r(4)\approx 0.4333,\ r(5)=0.2,\ r(6)=0$ for the case $m=6$ and $K>K_c$. Fig.\ref{fig4} shows the data for $r(A)$ meets the constrain as above mentioned. This again  tells one that the distribution function $f(\phi,\omega,t)$  can be used to obtain the parameter $r(A)$ through the symmetry transformation (\ref{trans}).

(e)When $A$ of the initial distribution $(0, A)$  is $\frac{2n\pi}m$ , the dynamics of the model is very sensitive to the initial conditions. The order parameters is defined as
\begin{eqnarray}r(A)= \frac1{\bar{n}}|\int_{-\infty}^{+\infty}\int_0^{2n\pi}f(\phi,\omega,t)e^{i\frac{\phi}m}d\omega d\phi +\Lambda|\label{partial order 2npidm}.\end{eqnarray}
The oscillators with initial phase very near  $\frac{2n\pi}m$ can either lag into the $n$-th cluster or drift forward to the $(n+1)$-th cluster. $\Lambda$ is the related order parameter for the entering  the $(n+1)$-th cluster, and the fraction of the oscillators for this kind is very sensive to the system's initial conditions. So do the parameters $\Lambda$ and $r(A)$.  $\bar{n}$ in Eq.(\ref{partial order 2npidm}) is in $(n, n+1)$ depending on parameter $\Lambda$. All are shown in Fig.\ref{fig5}.
\begin{center}
\begin{figure}[ht]
\begin{tabular}{cc}
\includegraphics[height=0.10\textheight, width=0.22\textwidth]{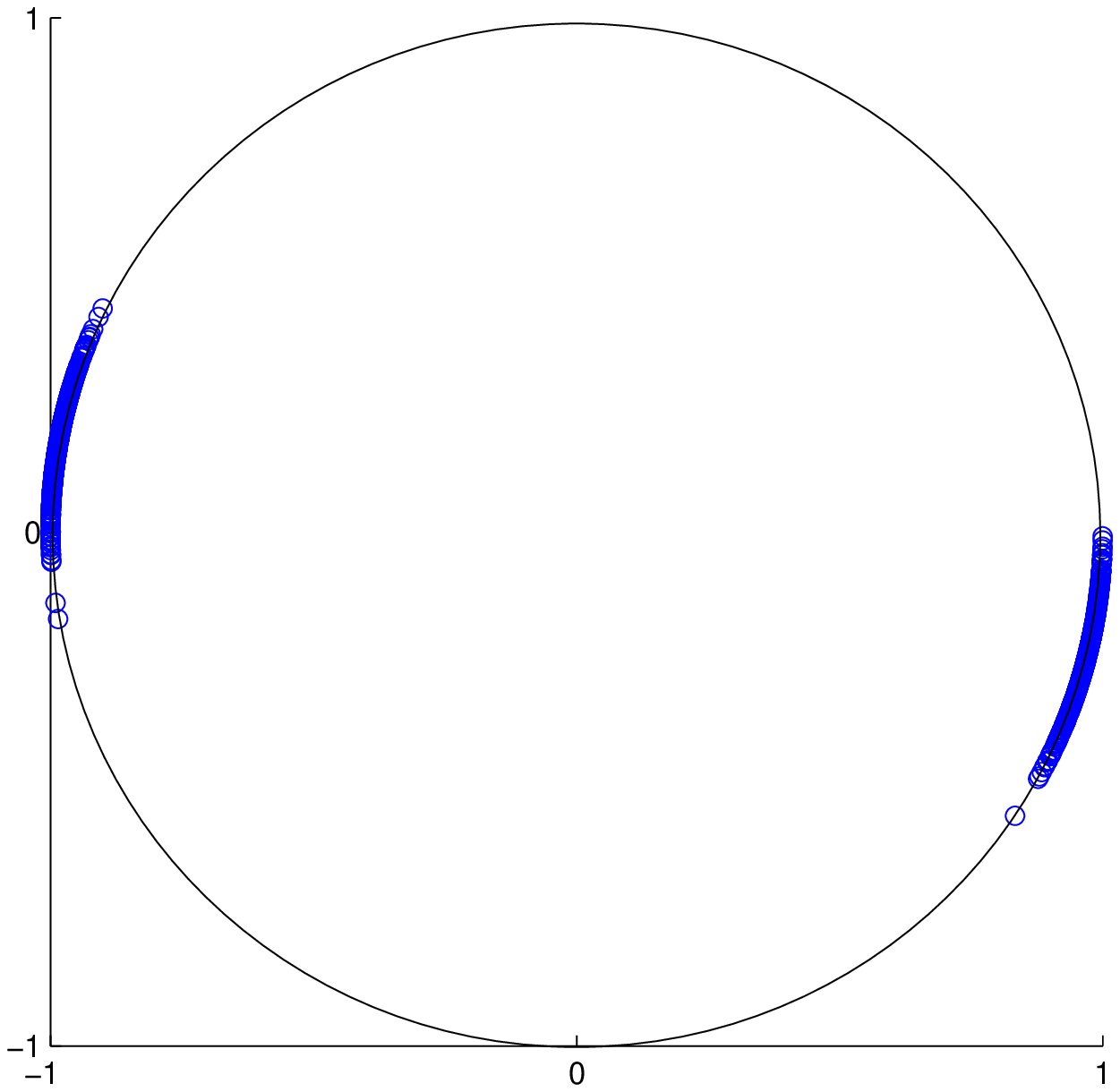}&
\includegraphics[height=0.10\textheight, width=0.22\textwidth]{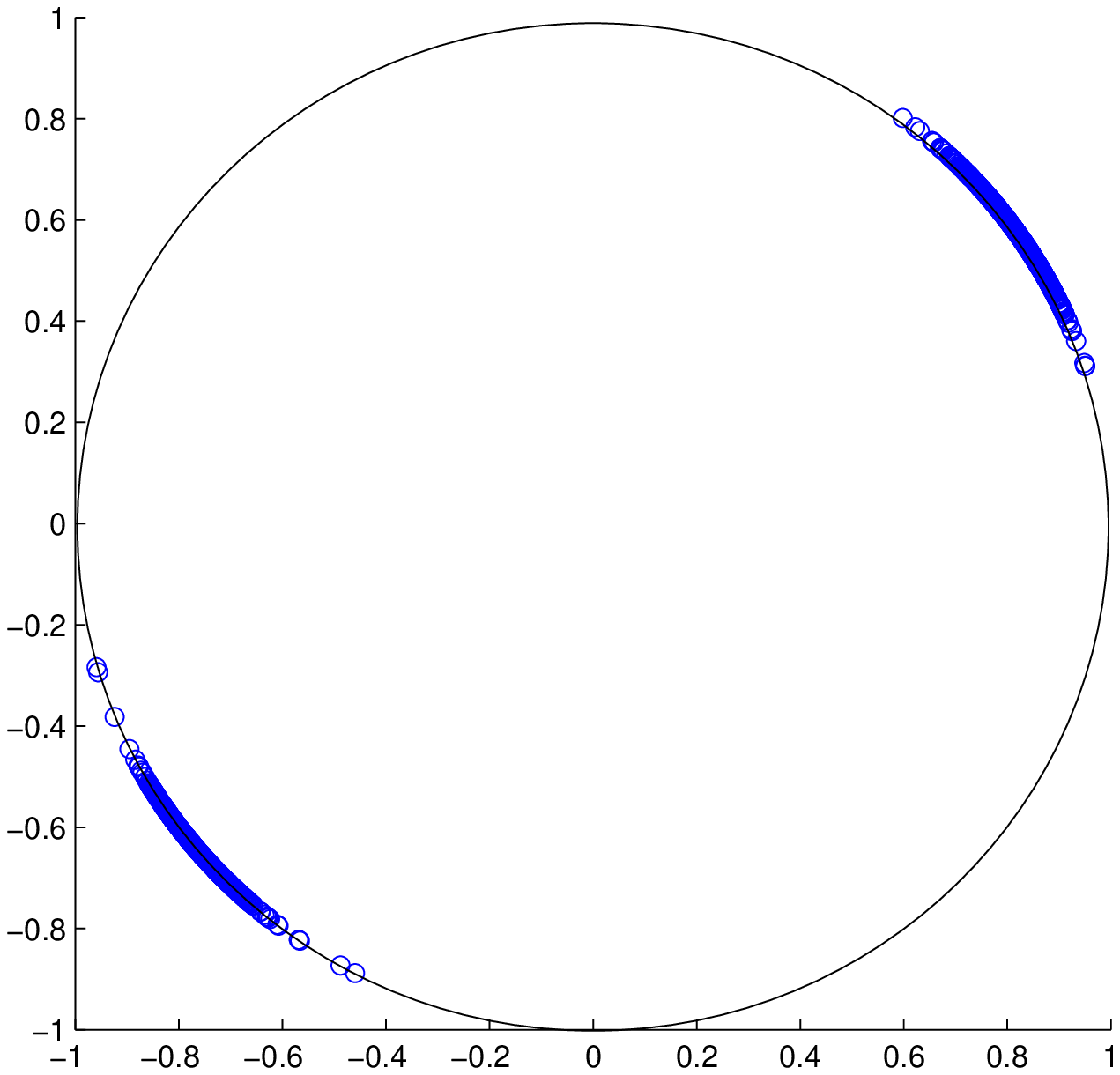}\\
\includegraphics[height=0.10\textheight, width=0.22\textwidth]{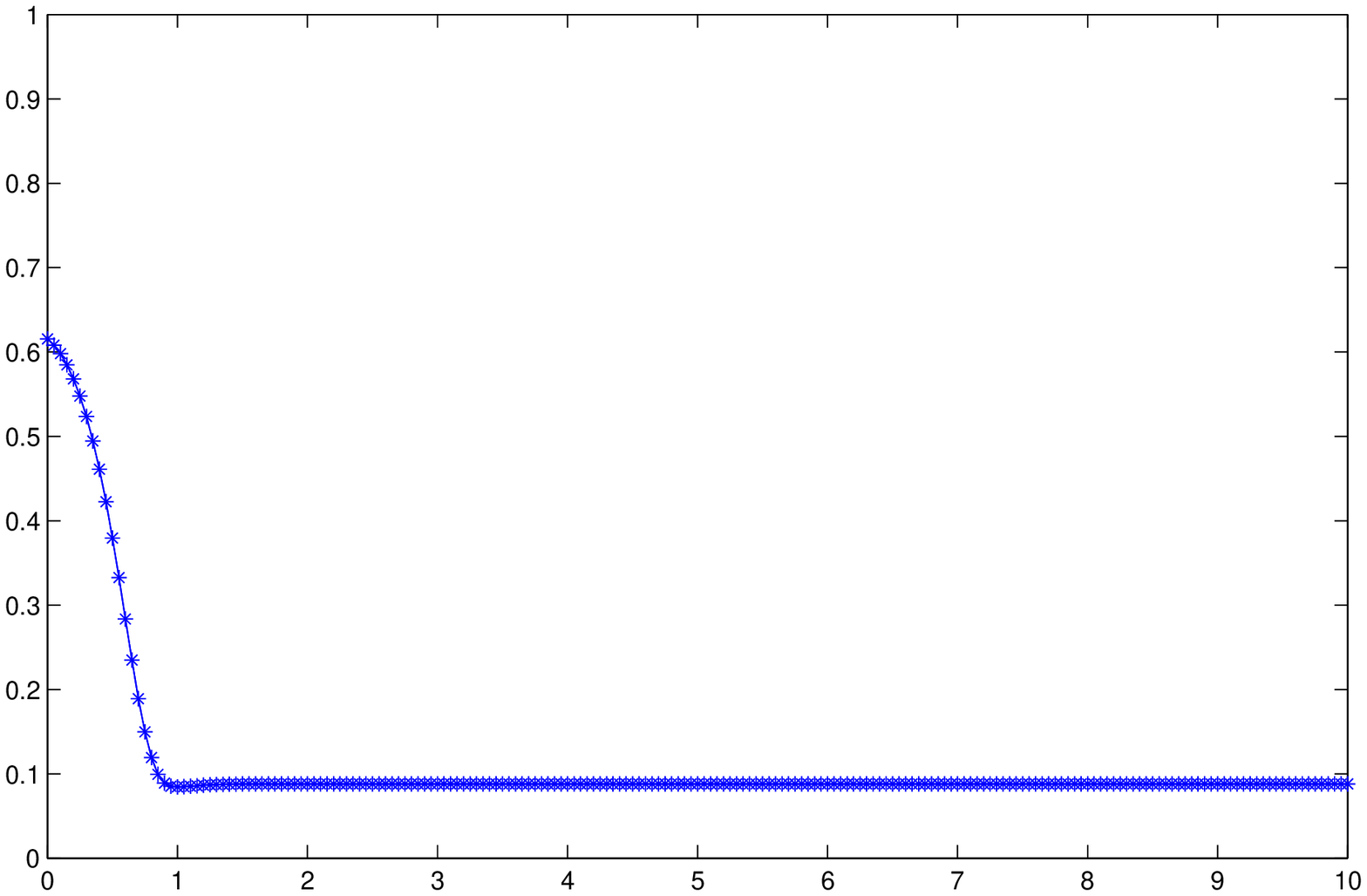}&
\includegraphics[height=0.10\textheight, width=0.22\textwidth]{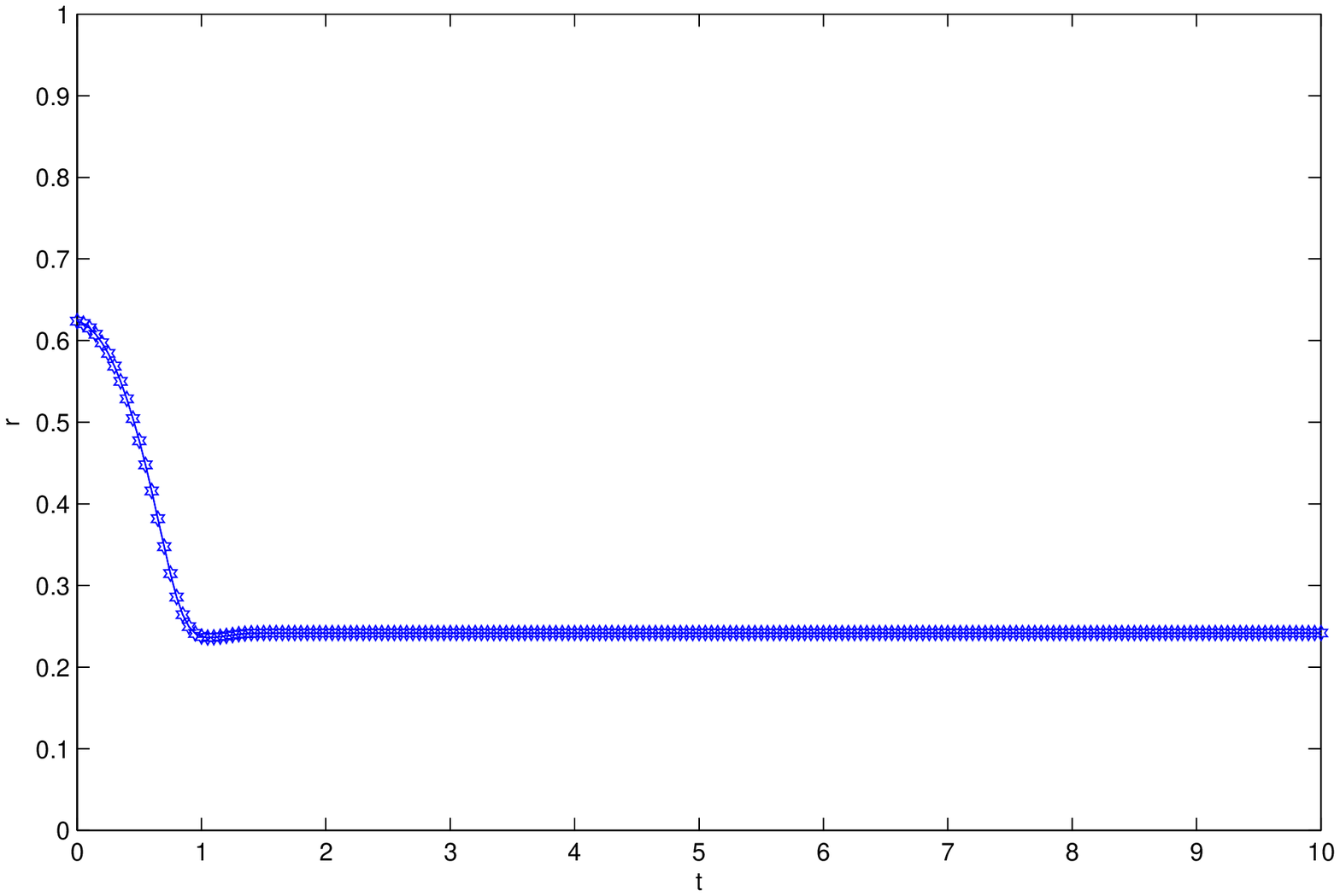}
\end{tabular}
\caption{Schematic diagrams $r(t)$ of initial uniform phases distribution in $(0, \pi)$   with the same parameters $K=5,\ m=2$. The only difference for the two cases is their initial values for the phases.} \label{fig5{fig4}}
\end{figure}
\end{center}

We now conclude the  symmetric viewpoint. The transformations (\ref{trans}) make it possible to relate the results of Eq.(\ref{Kuramoto 1-nth}) with that of Eq.(\ref{Kuramoto mth}). Furthermore, the periods   for both $m\theta$ in Eq.(\ref{Kuramoto mth}) and $\phi$ in Eq.(\ref{Kuramoto 1-nth}) are identical, that is, $2\pi$. The case $\Bar{K}_m>K_c>0$ in the model (\ref{Kuramoto mth}) is the same case $K>K_c$ for the transformed model (\ref{Kuramoto 1-nth}). So the initial phases distribution in $(0,A)$ in Eq.(\ref{Kuramoto mth}) is equivalent to the initial distribution $(0,mA)$ in Eq.(\ref{Kuramoto 1-nth}). So when $A\in (\frac{2n\pi}{m}, \frac{2(n+1)\pi}m)$, the initial phases distribution for variable $\phi$ is $(0,2(n+1)\pi)$. Every $2\pi$ distribution in Eq.(\ref{Kuramoto 1-nth}) will be a Kuramoto model and will be synchronized when $K>K_c$. Therefore the distribution $(0,2(n+1)\pi)$ for $\phi$ will be equivalent to $n+1$ Kuramoto models, that is, the $n+1$ cluster synchronization. This is the root of the formula (\ref{partial order}). Mathematically, the above results means that the phases lines $\frac{2n\pi}{m}, \ n=0,1,\cdots,\ m-1$ in the unit circle divide the interval and the switching across them is forbidden generally when $K>K_c$. Only the phases very very close to them could separate into either forward interval or backward interval due to their different frequencies. The forbidden lines give vivid demonstration of CS.

In addition, Eq.(\ref{Kuramoto mth}) is also invariant under the translation $\tilde{\theta}_n=\theta_n+\theta_0$, so CS phenomena is invariant under the translation of the initial conditions, see Fig.\ref{fig6} for details.
\begin{center}
\begin{figure}[ht]
\begin{tabular}{ccc}
\includegraphics[height=0.10\textheight, width=0.15\textwidth]{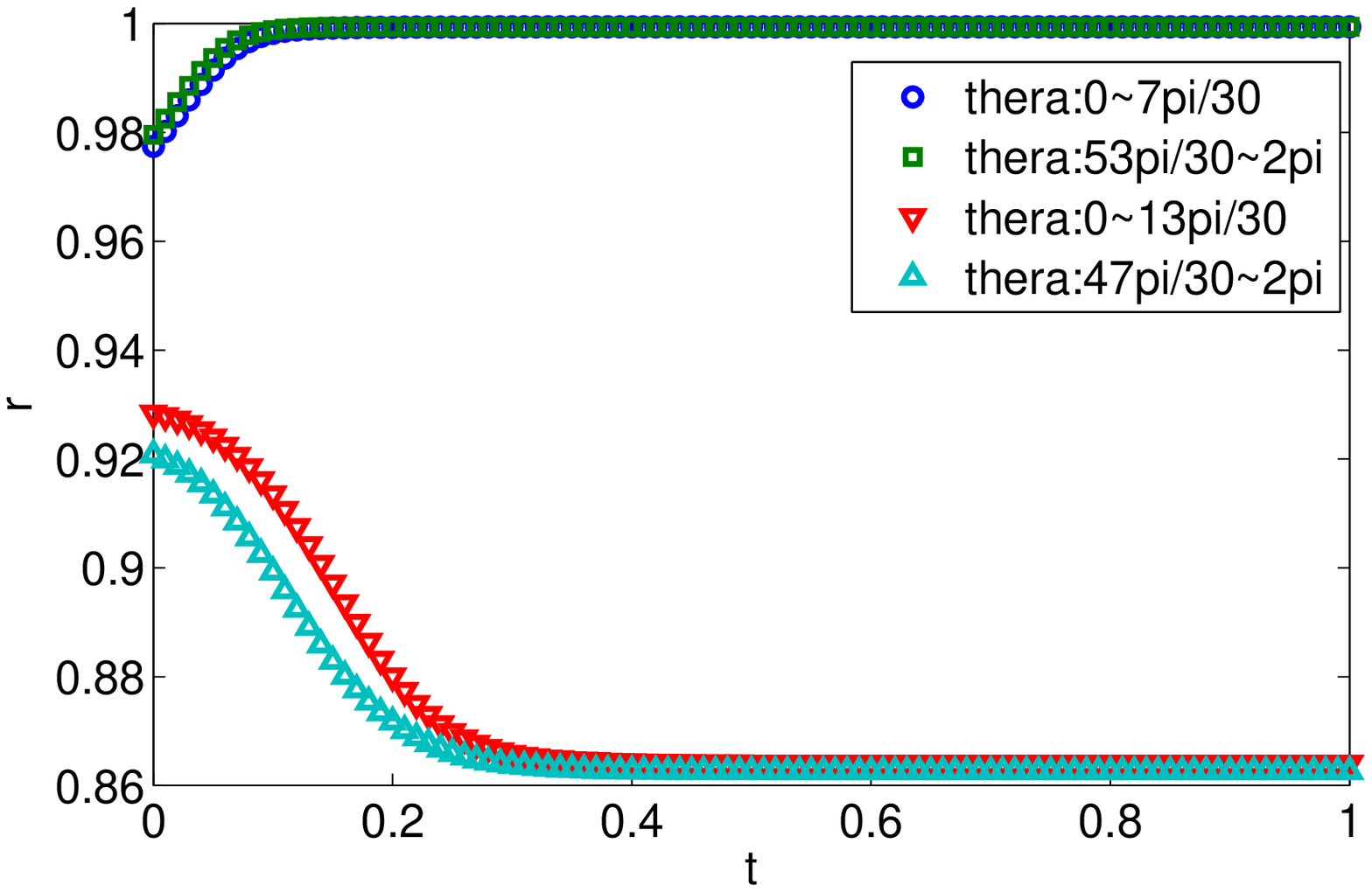}&
\includegraphics[height=0.10\textheight, width=0.15\textwidth]{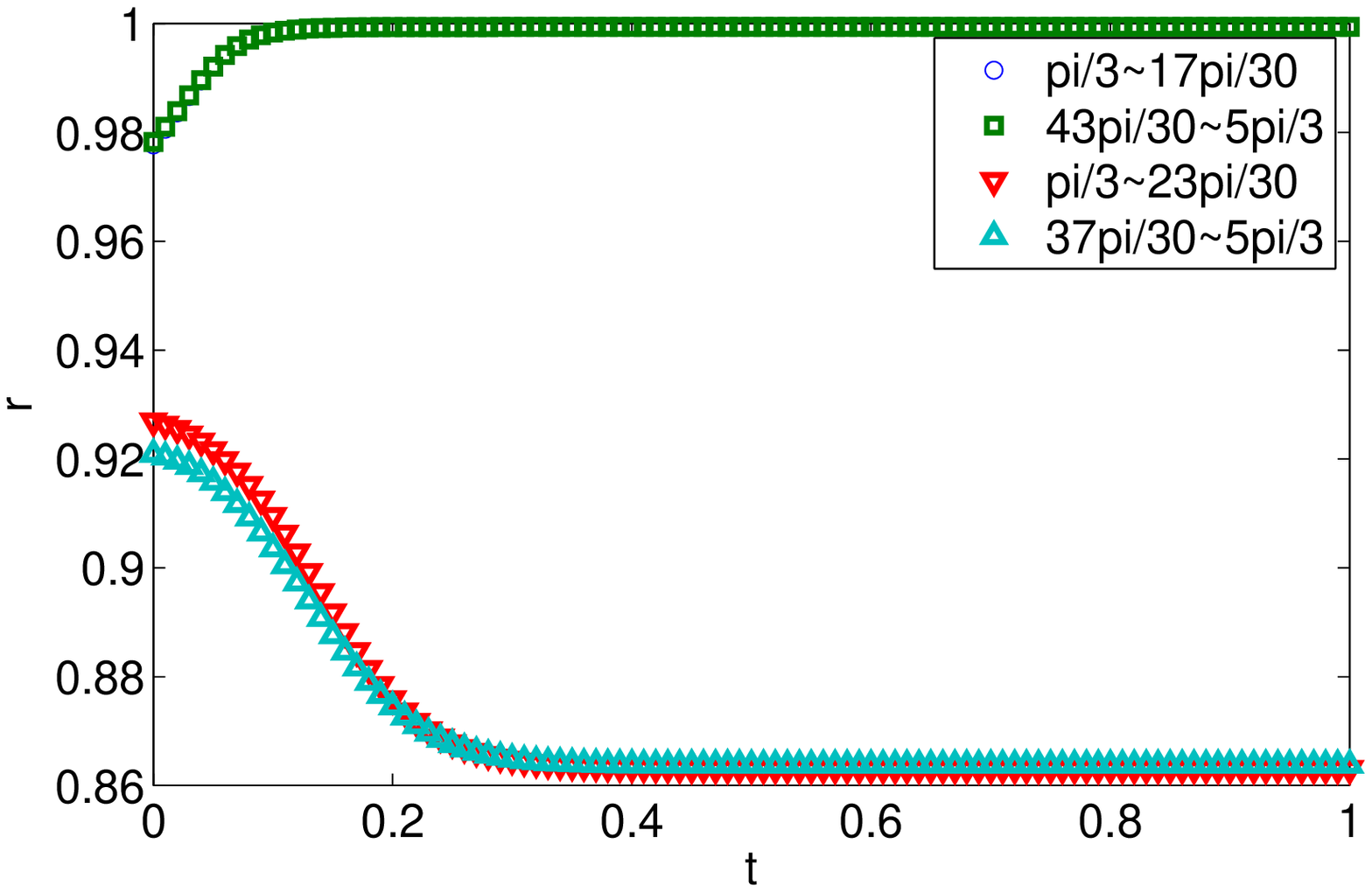}&
\includegraphics[height=0.10\textheight, width=0.15\textwidth]{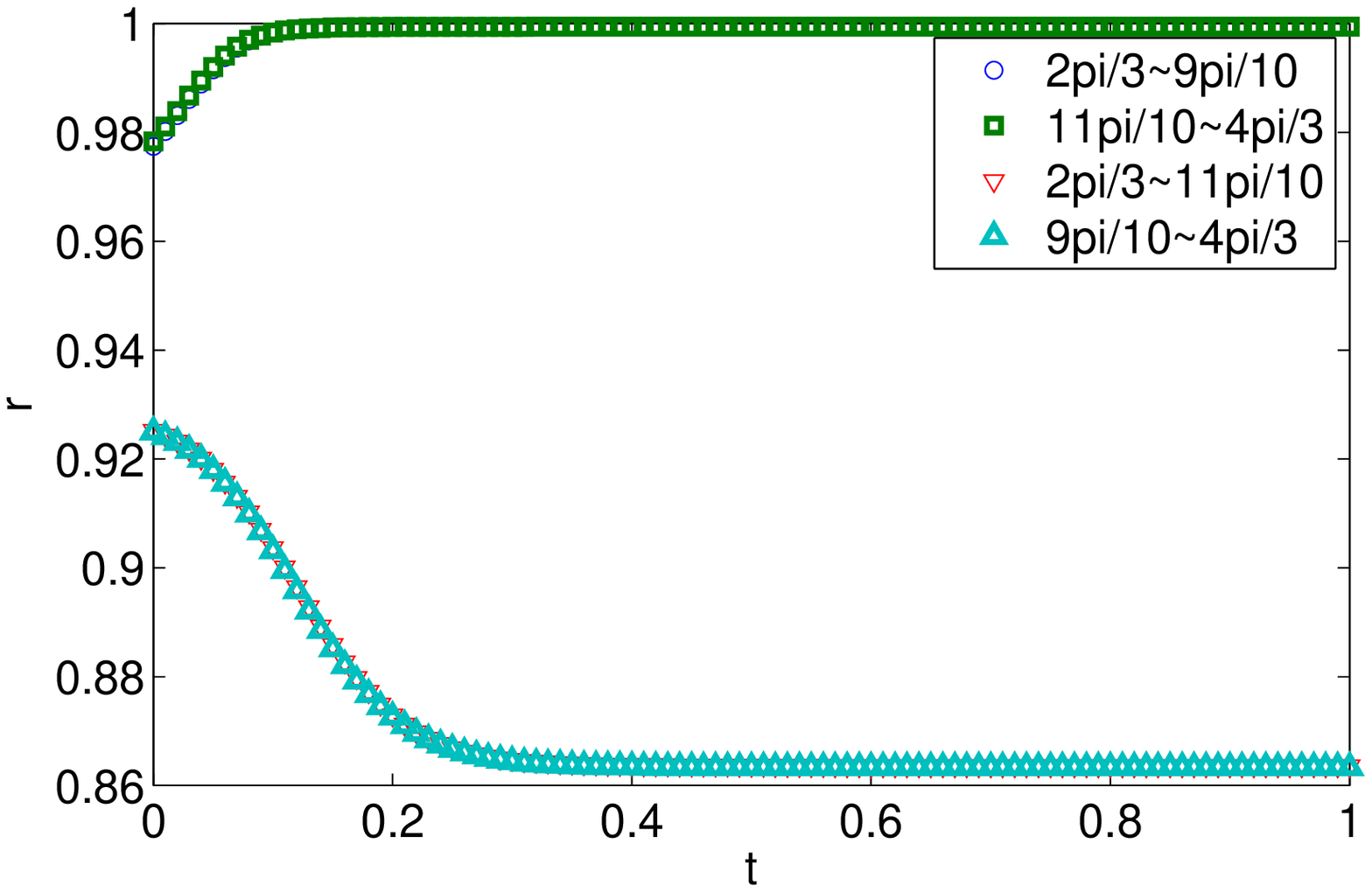}
\end{tabular}
\caption{Schematic demonstration of the translation properties for  $r(t)$ of initial uniform distribution of the phases in $(B,B+A)$ with the same coupling strength $K=5,\ K>K_c$ , the  same $m=6$ but different $B$.  }\label{fig6{fig5}}
\end{figure}
\end{center}
In the following, physical explanations are given on the basis of attraction and repulsion   interaction among the different oscillators .

The coupling strength $K>0$ in Eq.(\ref{Kuramoto ori}) is attracting to synchronization and is repulsive to synchronization as $K<0$. The incoherent state will stable in the case $K<0$.
Concerning the attractive and repulsive properties of the coupling parameter,
Hong and Strogatz have studied the identical oscillators with some couples others negatively (the contrarians) and some of positive coupling (the conformist). The contrarians like to be anti-phase with the mean field and the conformist is easy to in-phase. Also other interesting  phenomena like traveling wave occurs \cite{hong}. There also are references investigating the phenomena \cite{Freitas}-\cite{Pikovsky5}.
In Ref.\cite{Freitas}, the  $N$ identical phases with the non-linear coupling is studied and the positiveness and negativeness of the coupling parameter is controlled by the non-linearity coupling. The dynamics of the system is \begin{eqnarray}
&&\dot{\theta_i}=(1-\epsilon Z_i^2)\sum_{j=1}^{N_i}K_1\sin (\theta_j-\theta_i) \label{nonlinear Kuramoto},
\end{eqnarray}
where the local order parameter $Z_i=|\frac{1}{N_i}\sum_{j=1}^{N_i} e^{im\theta_j}|$. The repulsive coupling is realized if $1-\epsilon Z_i^2<0$. The nonlinear coupling will result in phase-locked states, while the large nonlinear coupling will give rise to multi-stable, periodic and chaotic states \cite{Freitas}. In the neural network, the fast studying model is attributed as $\dot{\theta_i}=\omega_i+\frac{1}{N}\sum_{j=1}^{N}K_{ij}\sin (\theta_j-\theta_i) $ with varying coupling strength as $K_{ij}=\alpha\cos(\theta_j-\theta_i) $.  This is actually the second Harmonic Kuramoto model. In this way,  the attractive and repulsive coupling parameter is achieved depending the difference of the two phases of the two oscillators \cite{Hansel1}-\cite{Hansel3}. Similarly, the attracting and repulsive properties of the coupling strength are the key to explain the cluster phenomena in the the higher odder harmonic coupling Kuramoto.

For the parameter $\Bar{K}_m>K_c>0$ , the coupling strength can be either attracting or repulsive
depending the difference of the two phases. If all $(\theta_j-\theta_i) $ are less than $\frac{2\pi}m$, then $m(\theta_j-\theta_i)<2\pi $ and they can be collectively synchronic and form a cluster and most oscillators synchronic in the cluster. Further increase of the range of the phases over $\frac{2\pi}m$, the oscillators with phases greater than $\frac{2\pi}m$ will be repulsed by the oscillators with phases less than $\frac{\pi}m$, so large phase difference will form for these two kinds of oscillators. Because the most oscillators cluster synchronically, the repulsion to the oscillators with phases larger than $\frac{2\pi}m$ is dominant, hence, these oscillators could only oscillators with phases large than $\frac{2\pi}m$ and the second cluster emerges. All along this way, more clusters will appear as the range of the phases becomes larger and larger.

Conclusion and discussion: in generalized Kuramoto with the higher order harmonic coupling, the view from the symmetry transformation  gives the explanation to CS both profound mathematical insight and   clear physical understanding. Detailed numerical studies confirm the symmetric analysis. The similar analysis could extend to the forced Kuramoto model $\dot{\theta_n}=\omega_n+\frac{1}{N}\sum_{j=1}^{N}K_m\sin m(\theta_j-\theta_n) +F_n(t) $, with the  force taking the form of $F_n(t)=F\sin\Omega t$ in the neural learning   network. Whenever the force is correlated with the oscillators, like $F_n(t)=F\sin(\Omega t-\theta_n)$, there is no symmetry group transformation like Eq.(\ref{trans}). Neither can the Kuramoto model with mixed higher harmonic orders coupling have symmetry group transformation like Eq.(\ref{trans}).  So new ideas are needed to be explored in the two cases.
\acknowledgments
The work was partly supported by the National Natural Science of China (No. 10875018)
and the Major State Basic Research Development Program of China (973 Program: No.2010CB923202).

\end{document}